\documentclass[sigconf]{acmart}

\usepackage[english]{babel}
\usepackage{blindtext}
\usepackage{graphicx}
\usepackage{subcaption}

\AtBeginDocument{%
  }

\setcopyright{acmlicensed}

\copyrightyear{2024}
\acmYear{2024}
\setcopyright{acmlicensed}\acmConference[HotOptics '24]{SIGCOMM Workshop on Hot Topics in Optical Technologies and Applications in Networking }{August 4--8, 2024}{Sydney, NSW, Australia}
\acmBooktitle{SIGCOMM Workshop on Hot Topics in Optical Technologies and Applications in Networking (HotOptics '24), August 4--8, 2024, Sydney, NSW, Australia}
\acmDOI{10.1145/3672201.3674119}        
\acmISBN{979-8-4007-0716-2/24/08}

\begin{document}

\title[OpticGAI]{OpticGAI: Generative AI-aided Deep Reinforcement Learning for Optical Networks Optimization}

\author{Siyuan Li}
\affiliation{
  \institution{\Large Shanghai Jiao Tong University}
  \city{Shanghai} 
  \country{China} 
}

\author{Xi Lin}
\authornote{Corresponding authors.}
\affiliation{
  \institution{\Large Shanghai Jiao Tong University}
  \city{Shanghai} 
  \country{China} 
}

\author{Yaju Liu}
\affiliation{
  \institution{\Large Shanghai Jiao Tong University}
  \city{Shanghai} 
  \country{China} 
}

\author{Gaolei Li}
\affiliation{
  \institution{\Large Shanghai Jiao Tong University}
  \city{Shanghai} 
  \country{China} 
}

\author{Jianhua Li}
\authornotemark[1]
\affiliation{
  \institution{\Large Shanghai Jiao Tong University}
  \city{Shanghai}
  \country{China} 
}

\begin{abstract}
    Deep Reinforcement Learning (DRL) is regarded as a promising tool for optical network optimization.
    However, the flexibility and efficiency of current DRL-based solutions for optical network optimization require further improvement.
    Currently, generative models have showcased their significant performance advantages across various domains.
    In this paper, we introduce OpticGAI, the AI-generated policy design paradigm for optical networks. 
    In detail, it is implemented as a novel DRL framework that utilizes generative models to learn the optimal policy network.
    Furthermore, we assess the performance of OpticGAI on two NP-hard optical network problems, Routing and Wavelength Assignment (RWA) and dynamic Routing, Modulation, and Spectrum Allocation (RMSA), to show the feasibility of the AI-generated policy paradigm. 
    Simulation results have shown that OpticGAI achieves the highest reward and the lowest blocking rate of both RWA and RMSA problems.
    OpticGAI poses a promising direction for future research on generative AI-enhanced flexible optical network optimization.
\end{abstract}



\maketitle

\section{Introduction}
Generative models, represented by the diffusion models~\cite{Rombach_2022_CVPR} and Large Language Models (LLM)~\cite{ChatGPT, GPT-4}, have exhibited remarkable performance across various fields.
This success is largely due to the powerful ability of generative models to extract high-quality features.
With their significant capacity for automatic content generation in diverse forms, it is evident that generative AI will profoundly impact numerous AI-assisted domains~\cite{du2024diffusion, li2024trustworthy}, including automated network optimization.

On the other hand, Deep Reinforcement Learning (DRL) has recently attracted widespread attention as a promising method in solving communication network problems~\cite{zhu2021network, gholami2022application}.
The most common application of DRL in network optimization is to use DRL agents to learn heuristic algorithms for sequential decision optimization~\cite{ge2023chroma, li2022digital}.
Nevertheless, this application still faces challenges such as requiring excessive computing resources and improving the flexibility of policy design~\cite{du2024diffusion, frohlich2021reinforcement}.

Recent research has also explored employing DRL to solve optical network optimization problems~\cite{Bernardez2021machine, li2020deepcoop}.
DRL agents can learn the open shortest path first weights to optimize the minimum maximum link load in networks~\cite{Bernardez2021machine}, or learn heuristic algorithms for dynamic routing, modulation, and spectrum allocation (RMSA)~\cite{Chen2018deep}, as well as for static routing and wavelength assignment (RWA)~\cite{cicco2022on}. 
On the other hand, researchers have attempted different advanced DRL algorithms for optical network optimization, such as Advantage Actor-Critic (A2C) for service provisioning in inter-domain Elastic Optical Networks (EON)~\cite{li2020deepcoop}, 
Asynchronous Advantage Actor-Critic (A3C) for RMSA in EONs~\cite{Chen2018deep}, and Multi-Agent Reinforcement Learning (MARL) for inter-domain RMSA~\cite{chen2019building}.
\begin{figure*}[!th]
    \centering
    \includegraphics[width=\linewidth]{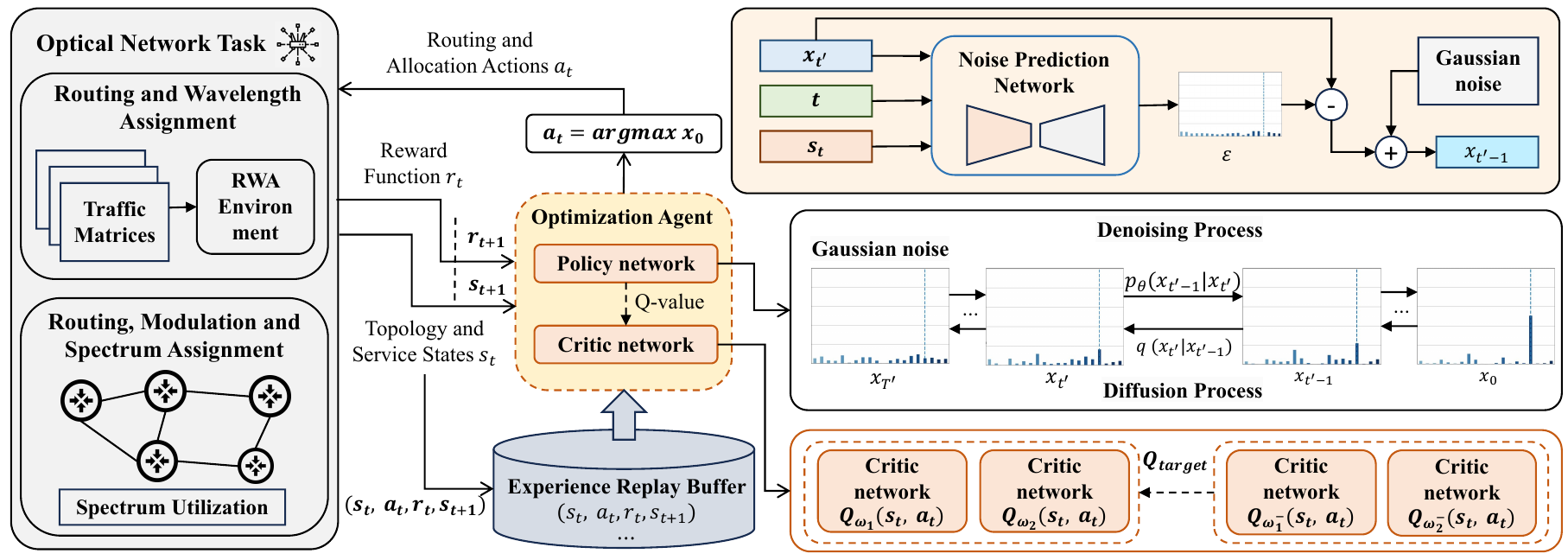}
    \caption{An instance of the OpticGAI paradigm: The diffusion-supported policy optimization algorithm for several optical network scenarios.}
    \label{overview}
\end{figure*}

However, practical applications of existing DRL-based optimization methods in optical networks still face several challenges:
(1) The wavelength assignment problem needs to be considered to ensure there are no wavelength conflicts, while the scheduling and path selection of optical signals also need to be considered;
(2) Optical networks typically use complex topological structures, which poses a significant challenge for learning algorithms to understand and make accurate optimization decisions.
(3) Dynamic changes in optical network traffic require optimization algorithms to respond quickly and efficiently.
However, existing DRL methods lack the flexibility to adapt to complex and dynamic optical network environments, which limits their further performance improvement.
This is largely due to the difficulty of fixed neural network structures to learn different network conditions and service states of optical networks.
These challenges pose an urgent need for more flexible learning algorithms for optimization policy.
Current DRL algorithms often generate unimodal policy distributions~\cite{liu2024multitask, du2024diffusion}, potentially leading to local optima and poor adaptability in complex optical network environments~\cite{yang2023policy}. 
In contrast, generative models are good at capturing complex probability distributions~\cite{Rombach_2022_CVPR, wang2022diffusion}.
In this work, we ask: Is it possible to utilize the powerful generation capability of AI-Generated Content (AIGC) technology to improve the flexibility and performance of existing DRL-based optical network optimization algorithms?

To address these challenges, we propose OpticGAI, an AI-generated policy paradigm that considers the search capability of AIGC-enabled reinforcement learning for optical network optimization~\cite{liu2024multitask, du2024diffusion}.
In this work, we propose a diffusion model-based policy network as a case to address the challenges of DRL-based optimization in optical networks.
In detail, OpticGAI integrates the diffusion model with the utility function of the optical network to form a policy network, capture complex relationships between states and decisions, and adapt to dynamic and complex optical network environments.
The reverse process of the Denoising Diffusion Probability Model (DDPM) ~\cite{ho2020denoising} serves as the policy network for service selection schemes in optical networks.
Furthermore, we take two fundamental problems, RWA and RMSA, as case studies to show how AIGC can be employed to improve DRL for network optimization further.
In addition, we investigate the learning ability of OpticGAI in both tasks compared with existing advanced algorithms.
The advantages of OpticGAI are demonstrated in terms of network blocking rate, outperforming all compared heuristic algorithms and DRL-based algorithms in both tasks. 
In conclusion, our contributions are summarized as follows:
\begin{itemize}
    \item We propose OpticGAI, an AI-generated policy design paradigm for optical networks that utilizes generative models to learn flexible policy networks instead of fixed ones.
    \item We develop a diffusion-supported policy optimization mechanism to capture complex and dynamic features of several optical network environments.
    \item We analyze our findings and discuss the challenges and future opportunities brought by the AIGC-enhanced optical network optimization idea.
\end{itemize}

The rest of this paper is organized as follows. 
Section 2 introduces the framework of OpticGAI, followed by the detailed diffusion-generated policy and the generated policy for optical network optimization.
Then we evaluate the performance of OpticGAI about achieved reward and the blocking rate on two optical network problems, RWA and RMSA in Section 3.
In Section 4, discussions and challenges about the generalization capabilities and explainability of OpticGAI are introduced.
Besides, we introduce some related works about DRL algorithms for optical network optimization tasks and propose our analysis of these algorithms in Section 5. 
Finally, we conclude this paper in Section 6.

\section{AI-Generated Policy Design Paradigm for Optical Networks}
\subsection{Structure of OpticGAI}
As mentioned earlier, DRL algorithms often tend to generate unimodal policy distributions, which may lead to convergence to local optima and hinder adaptation to changes in the optical network environment and complex scenes~\cite{yang2023policy}. 
OpticGAI employs generative models instead of traditional policy networks to capture complex policy distributions and adapt to changes in complex scenes of optical networks.
Compared to traditional DRL algorithms that tend to generate unimodal policy distributions, OpticGAI largely avoids the situation of convergence to local optima.
It models the utility function of the optical network and integrates a generative model for generating a policy network, to learn the complex and dynamic optical network environment and effectively utilize the resources.

This paper introduces an instance of the OpticGAI paradigm, the diffusion-supported policy optimization, which is illustrated in \autoref{overview}.
Employed as the policy network in DRL algorithms, the diffusion model can better balance exploration and exploitation. 
By capturing the mapping relationship between states and decisions, the diffusion model ultimately generates actions, namely the optical network optimization decision scheme.
Let $t$ and $\tau$ represent the steps in the diffusion process and the Markov decision process, respectively.
We adopt the reverse process of DDPM as the policy network $\pi_\theta(s_{\tau})$, as shown in \autoref{overview}.
The policy network samples noise from a Gaussian distribution in time step $\tau$.
The noise prediction network predicts the noise to be removed at denoise step $t$ based on the state $s_{\tau}$ and the action at the last denoise step $a_{t-1}$.
After $T$ denoise steps, the optimization scheme is generated.

\subsection{Diffusion-Generated Policy}
The diffusion model captures complex probability distributions through learning to denoise random noise and processes them through multiple denoising processes to generate content~\cite{Rombach_2022_CVPR}.
When the diffusion model is used as a policy network in the online DRL algorithm, the input of the noise prediction network includes the network and service states.
We use the DDPM to introduce how diffusion models work as policy networks.
DDPM has two opposite processes, the diffusion process and the denoising process.
One of them is the diffusion process $q(A_t | A_{t-1})$, which refers to adding Gaussian noise to scheme $A_0 \sim q (A_0)$ during $T$ denoise steps, gradually transforming $A_0$ into Gaussian noise $A_T$:
\begin{equation}
    q(A_{1: T} \mid A_0)
    = \prod_{t=1}^T \mathcal{N}( A_t ; \sqrt{1-\beta_t} A_{t-1}, \beta_t \mathbf{I} ).
\end{equation}
The other, the denoising process, involves a $T$ step denoising operation on Gaussian noise $A_T \sim \mathcal{N} (0, \mathbf{I})$ to return to $A_0$~\cite{ho2020denoising}:
\begin{equation}
    p_\theta(A_{0:T}) 
    = p(A_T)\prod_{t=1}^T \mathcal{N}(A_{t-1} ; \mu_\theta(A_t, s_{\tau}), \Sigma_\theta(A_t, s_{\tau})).
\end{equation}
For the optical network optimization problem, such as RWA, building expert scheme sets is often impossible, so we ignore the forward process.
Instead, we focus on the reverse process of the DDPM which is used as the policy network, as shown in \autoref{overview}. 

The policy network samples noise from the Gaussian distribution at time step $\tau$. 
The noise prediction network predicts the noise to be removed at denoise step $t$ based on the state $s_{\tau}$ and the output $A_{t-1}$ from the previous time step. 
After performing denoising operations over $T$ steps, the scheme $A_0$ of optical network problems is generated. 
The action $a_{\tau}$ is then determined as the decision scheme corresponding to the maximum probability in the vector $A_0$.

\subsection{Generated Policy for Optical Network Optimization}
In addition to policy networks, effective state representation and action space are crucial for the learning process.
Large observations bring scalability and learning challenges, while larger action spaces increase the complexity of the learning phase. 

Taking the RWA problem as an example, our goal is to use the trained OpticGAI agent as a specific network input, represented by a given traffic matrix, to create a specialized heuristic method and output feasible solutions~\cite{cicco2022on}. 
Let $E$, $W$, and $P$ denote the set of bidirectional links, set of available wavelengths in each link, and set of pre-computed $K$-shortest paths, respectively. 
Then static RWA problem can be formalized as the ILP formula~\cite{ozdaglar2003routing}, the objective is maximizing 
\begin{equation}
    \mathcal{L} =  \sum_{p \in P} \sum_{w \in W} x_p^w,
    \label{equ1}
\end{equation}
where $x_p^w \in\{0,1\}$, with the constraints of $\sum\limits_{p \mid l \in p} x_p^w \leq 1, \forall l \in E$ and $\sum\limits_{p \in P_{(s, d)}} \sum\limits_{w \in W} x_p^w \leq \rho_{(s, d)}$, where $s$, $d$ are two different nodes.
An event corresponds to an RWA problem instance with uniform traffic distribution.
In each event, the agent decides whether to accept or reject a connection request based on observations.
At the end of an episode, resources are reset, and a new traffic matrix is processed.
The observation space of OpticGAI includes the source-destination pair of the current request and the spectrum occupancy of the $K$ shortest paths, forming a vector of dimension $(2|V| + 2K)$, where $V$ represents the set of routing nodes.
OpticGAI generates a discrete action space with a dimension of $K + 1$.
In each training iteration, OpticGAI collects transformations from the interaction between the environment and policy agent and samples them in subsequent iterations to update the parameters of the critical network and policy network until convergence, thereby generating the optimization algorithm.

For the RMSA task, we maintain consistency with~\cite{natalino2020optical, Chen2018deep}, utilizing the similar Markov decision process model for OpticGAI.
we adopt a paradigm where a remote SDN controller interacts with local SDN agents to gather network state and lightpath requests and to distribute RMSA solutions~\cite{Chen2018deep}. 
Upon receiving a lightpath request $R_{\tau}$, the SDN controller retrieves key network state representations from the traffic engineering database, including active lightpaths, resource utilization, and topology, to generate the state $s_{\tau}$.
The policy network of OpticGAI processes the state information and outputs the RMSA policy $\pi_\theta(A_{\tau}|s_{\tau}, \theta_\pi)$, where $A_{\tau}$ is the set of candidate RMSA solutions and $\theta$ represents the parameters of the policy network.
The controller then acts on $\pi_{\tau}$ by selecting action $a_{\tau} \in A_{\tau}$ and attempts to establish the corresponding lightpath.
Then the environment receives the outcomes related to the previous RMSA operation as feedback and generates an immediate reward $r_{\tau}$ for OpticGAI. 
The tuple $(s_{\tau}, a_{\tau}, r_{\tau})$ is stored in the experience buffer, from which OpticGAI derives training signals for subsequent updates of the generation network. 
The objective of OpticGAI while servicing requests is to maximize long-term cumulative rewards.
Ultimately, OpticGAI can learn adaptable RMSA policies through dynamic lightpath provisioning.

\section{Evaluation}
In this section, we evaluate the performance of our proposed OpticGAI framework on representative optical network optimization problems to show its feasibility.
We implement our OpticGAI approach based on the Optical RL-Gym~\cite{natalino2020optical}, an open-source toolkit for applying RL to optimization problems in optical networks. 
As proof-of-concept experiments, we evaluate the performance of OpticGAI on RWA and RMSA tasks.

\subsection{Evaluation on RWA Task}
\begin{figure}[!t]%
    \centering
    \subfloat[10-node network topology for RWA task.]{
        \includegraphics[width=0.43\linewidth]{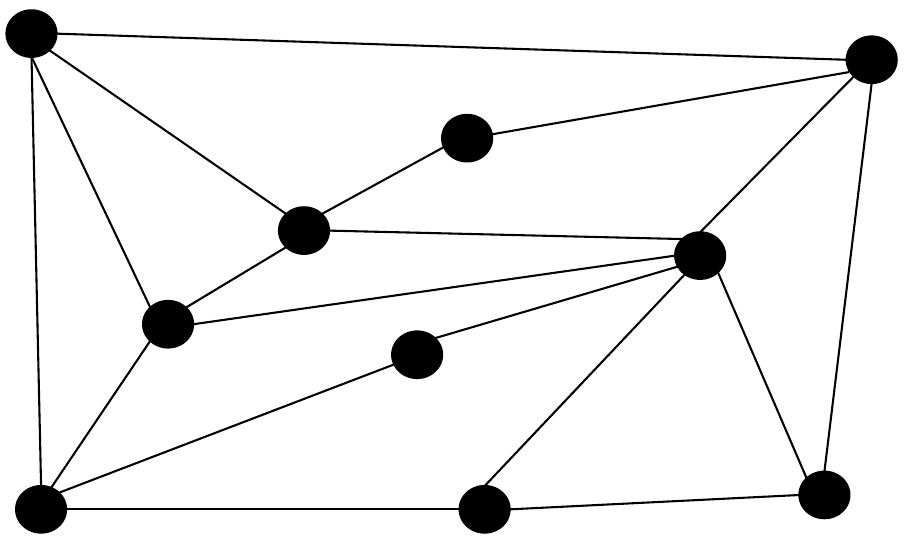}
        \label{fig-topo-1}
    } 
    \hspace{0.02\linewidth}
    \subfloat[14-node network topology for RMSA task.]{
        \includegraphics[width=0.43\linewidth]{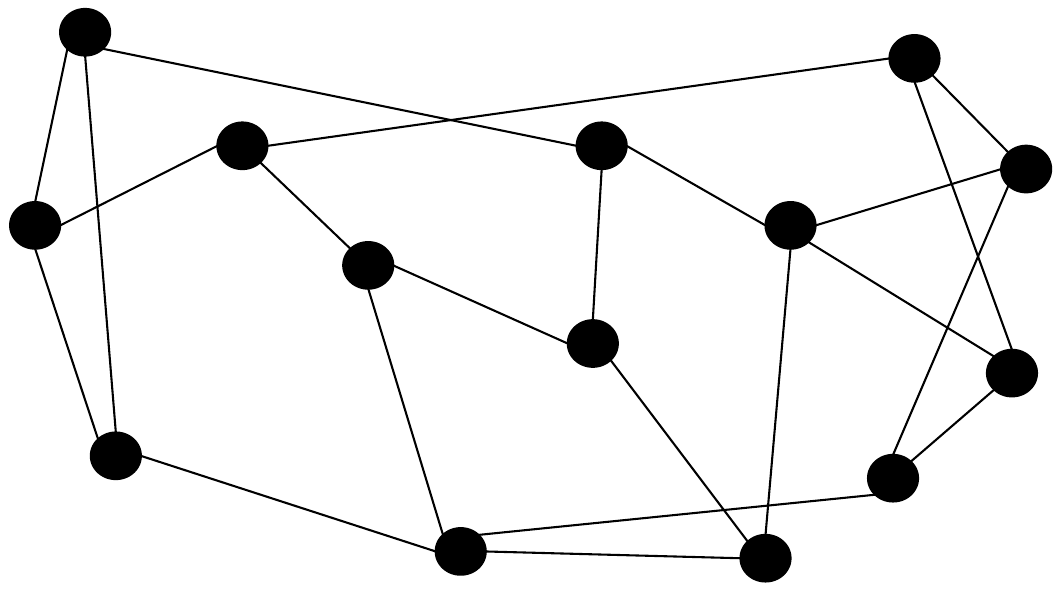}
        \label{fig-topo-2}
    }
    \caption{Network topology for RWA and RMSA tasks respectively.}
    \label{topo}
\end{figure}

\textbf{Setup.}
The RWA problem is to allocate routing and wavelength for a group of optical path requests in an optical wavelength division multiplex network.
Our goal is to create a specialized heuristic method for the RWA problem under the assumption of wavelength continuity constraints applied to each optical path, outputting feasible solutions from a given flow matrix.
As shown in \autoref{topo}, we first evaluate the performance of OpticGAI on the static RWA problem of a 10-node optical network. 
Following~\cite{cicco2022on}, we simulate using a representative 10-node reference network and employ the sparse rewards and shaped rewards to efficiently learn in networks with high link capacity.
We use the Proximal Policy Optimization (PPO) algorithm implemented as PPO-Full, and PPO-FF with first-time fitting policy as the DRL baselines, together with Integer Linear Programming (ILP) for the optimal solution. 
In addition, we also implement the heuristic algorithms Shortest Path fitting (SP), K-Shortest Path Fitting (K-SP), Least Loaded Path (LLP), and the optimality bounds of ILP as the baselines.
\begin{figure}[!t]
    \centering
    \includegraphics[width=\linewidth]{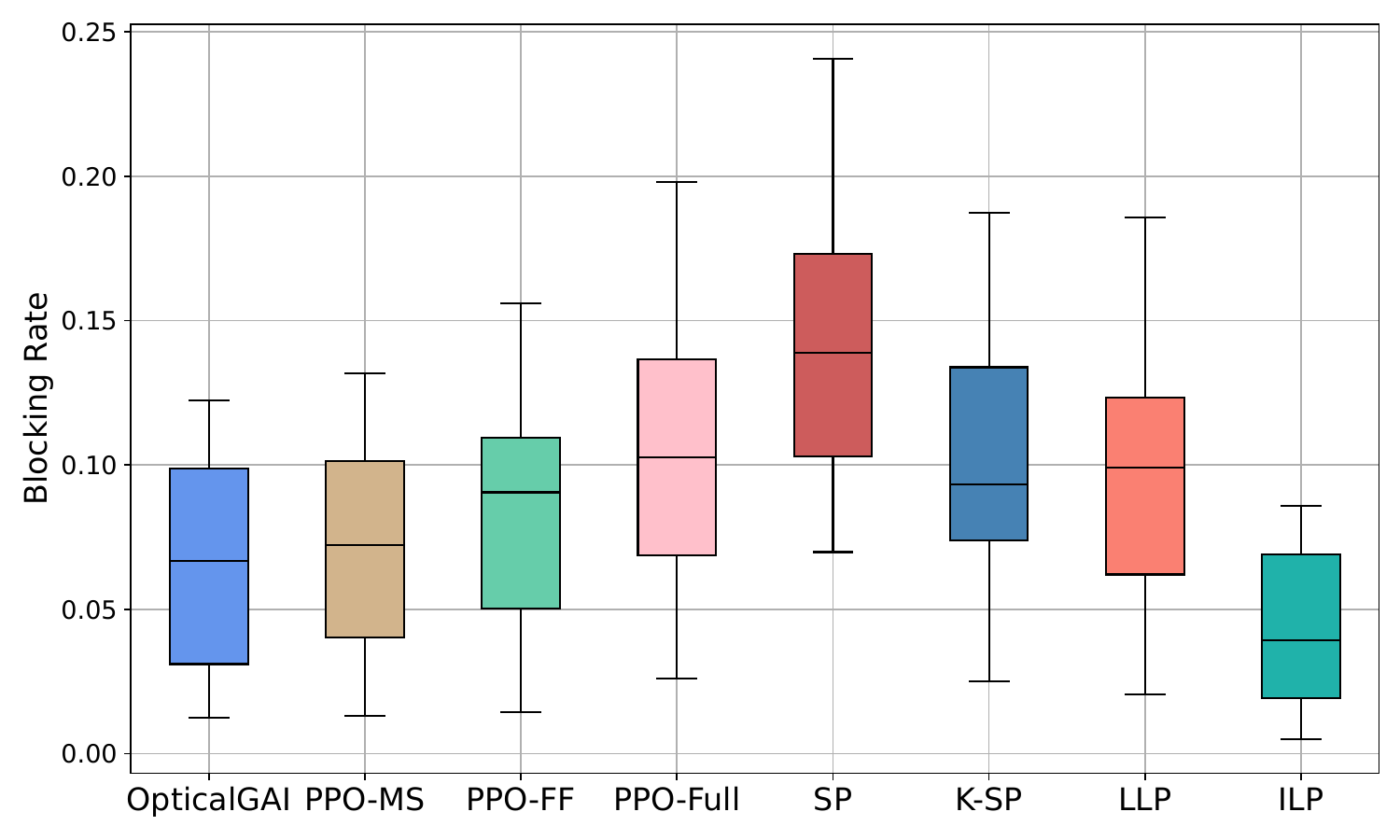}
    \caption{Blocking rate of OpticGAI, PPO, heuristic algorithms, and ILP on the RWA task.
    }
    \label{RWA-BlockingRate}
\end{figure}

\textbf{Baselines.}
For the RWA problem, we compare the proposed OpticGAI method with some heuristic algorithms and DRL methods:
Firstly, heuristic algorithms include greedy algorithms implemented as SP and K-SP, as well as LLP algorithms with minimum load.
Secondly, advanced PPO reinforcement learning algorithms are implemented using the first fitting PPO-FF algorithm, the PPO-Full algorithm that can handle any wavelength available in $K$ paths, and the PPO-MS algorithm that adopts the multi-start strategy.
Finally, the commercial solver Gurobi \footnote{https://www.gurobi.com} is used to solve the ILP formula as \autoref{equ1}, obtaining an accurate solution.
Other experimental settings are following~\cite{cicco2022on}.
\begin{figure}[!t]
    \centering
    \includegraphics[width=\linewidth]{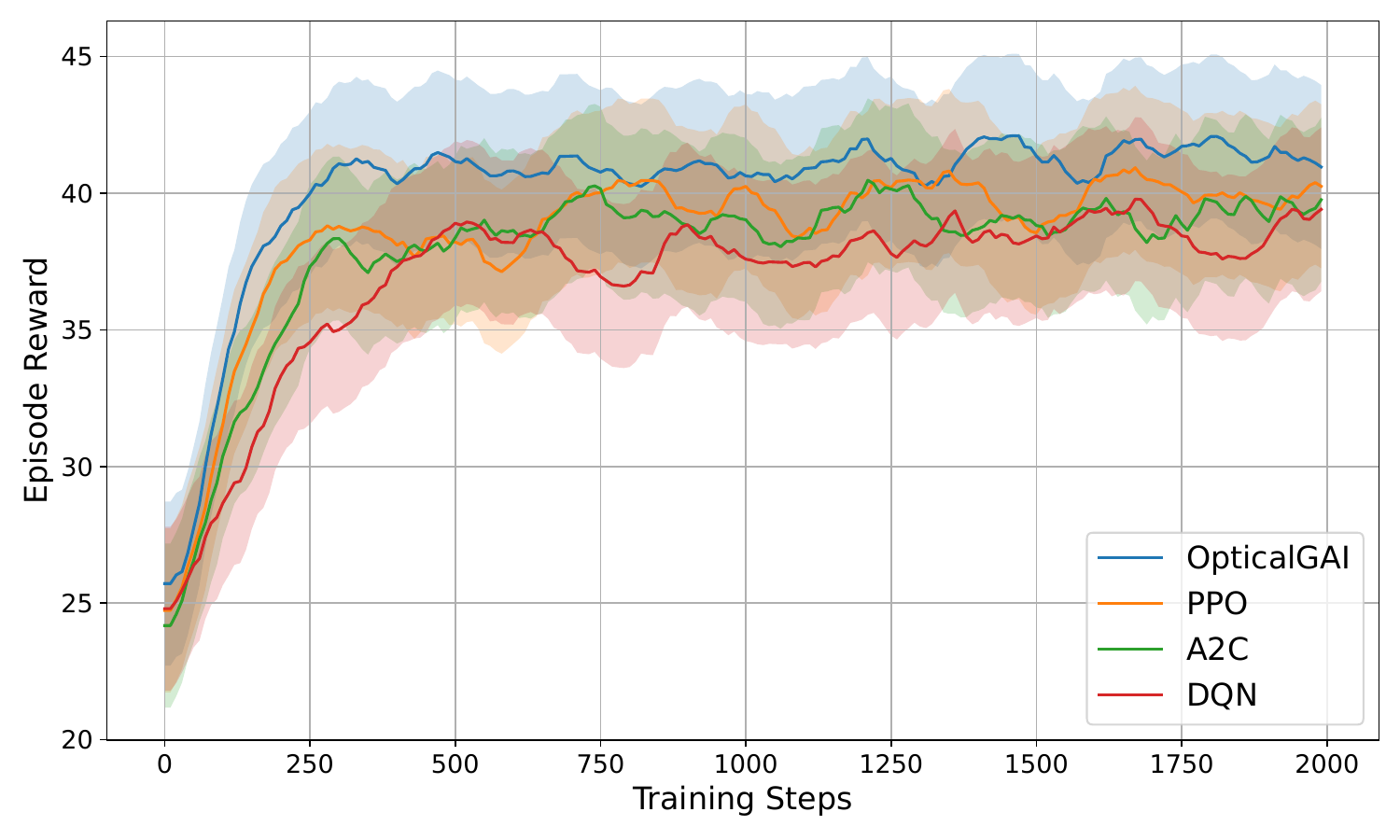}
    \caption{Episode reward during training on the RMSA task using OpticGAI and other DRL agents.
    }
    \label{RMSA}
\end{figure}
\begin{figure}[!t]
    \centering
    \includegraphics[width=\linewidth]{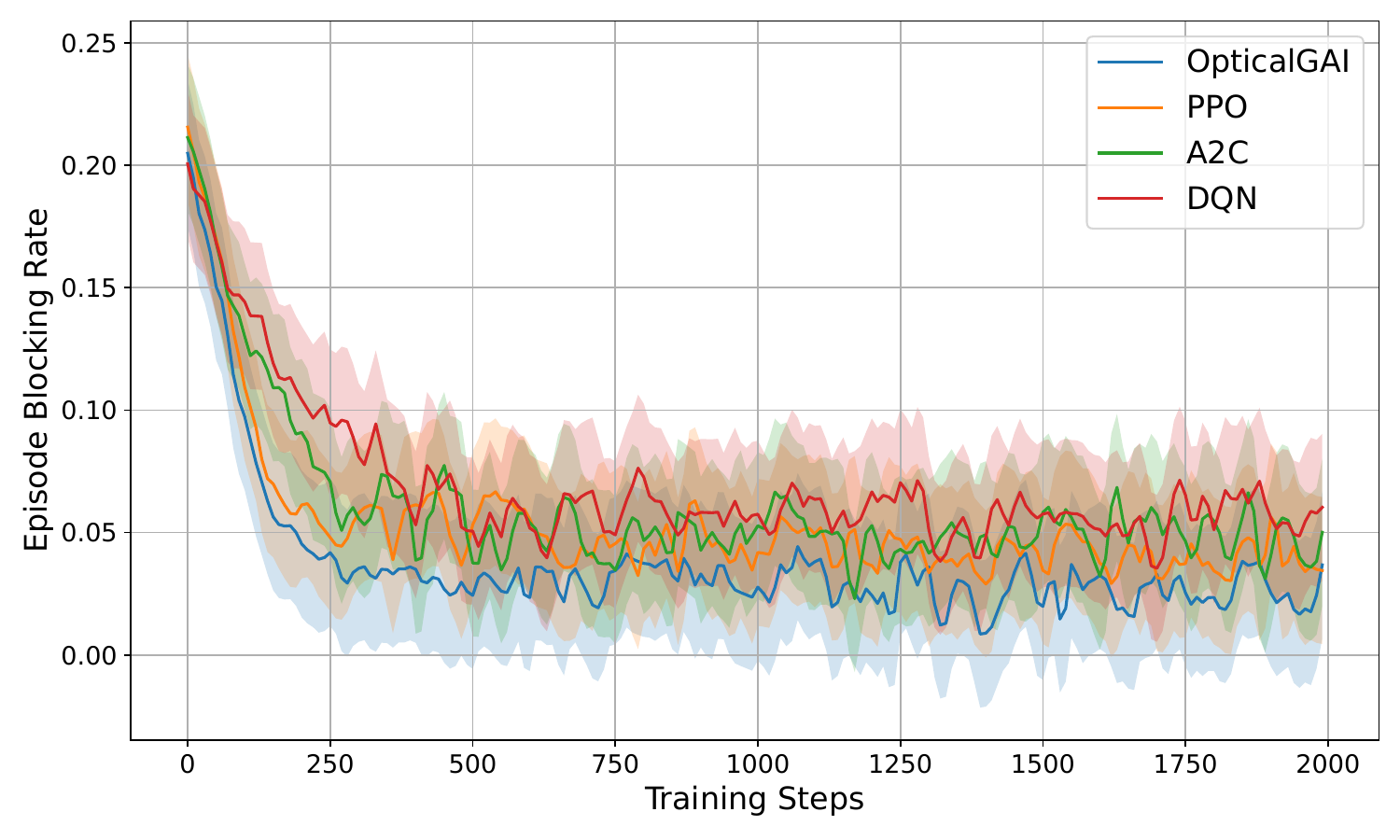}
    \caption{Episode blocking ratio during training on the RMSA task using OpticGAI and other DRL agents.
    }
    \label{RMSA-BlockingRate}
\end{figure}

\textbf{Results.}
We evaluate the blocking rate of OpticGAI, advanced PPO, heuristic algorithms, and ILP on 500 traffic matrices of RWA tasks.
The blocking rate is a critical performance metric in optical network optimization, indicating the proportion of connection requests that cannot be accommodated due to resource limitations.
As illustrated in \autoref{RWA-BlockingRate}, OpticGAI achieves the lowest blocking rate, indicating superior performance in optimizing optical network resources compared to existing optimization algorithms.
PPO variants (PPO-MS, PPO-FF, PPO-Full) show varying degrees of effectiveness, with PPO-MS and PPO-Full performing better than PPO-FF.
Heuristic algorithms like SP and K-SP exhibit higher blocking rates, indicating less efficiency in wavelength resource allocation.
LLP performs moderately well, balancing load across the network to reduce the blocking rate but not as effectively as OpticGAI.
While ILP shows precise effectiveness with a low blocking rate, 
its lack of scalability due to the NP-hard nature of the RWA problem limits its practical applicability.
ILP's high computational complexity makes it impractical for larger-scale problems, serving primarily as a benchmark.
OpticGAI's superior performance underscores its potential for real-world optical network optimization.

\subsection{Evaluation on RMSA Task}
\textbf{Setup.}
In EON, the resource allocation problem is commonly referred to as the RMSA task, which we implement using the \textit{DeepRMSAEnv} environment provided in Optical RL-Gym, which implements specific state representations, action mappings, and reward functions for dynamic RMSA use cases. 
In addition to the static RWA task, we also evaluate the dynamic RMSA task in EON on NSFnet topology using the \textit{DeepRMSAEnv} environment provided in Optical RL-Gym. 
We employ the same representative 14-node NSFNET topology as the official implementation, and each fiber link can accommodate 100 frequency slots.
For service requests, we assume that they generate a Poisson distribution that follows a uniform traffic distribution.
The required bandwidth for each request is evenly distributed between 25 and 100 Gb/s.
The average arrival interval and service duration interval of the service are 10 and 25 time steps, respectively.
All other parameters related to reinforcement learning remain the default.

\textbf{Baselines.}
For the RMSA problem, we compare the proposed OpticGAI method with four advanced DRL methods, Deep Q-Network (DQN)~\cite{DQN}, A2C~\cite{A2C}, and PPO~\cite{PPO}, demonstrating the superior performance of OpticGAI. 
We use the algorithm provided by the stable baseline for implementation\footnote{https://github.com/hill-a/stable-baselines}, and the parameter settings follow~\cite{natalino2020optical}.
The learning rate is set to $10^{-5}$, and the employed neural network has 4 layers, each with 150 neurons, and all other parameters are set to default values.

\textbf{Results.}
To maintain consistency, the experiment is conducted on NSFnet topology with~\cite{Chen2018deep, natalino2020optical}, Under the same modulation format, service intensity, number of available spectrum blocks, and set length.
As shown in \autoref{RMSA}, the OpticGAI model is the first model to achieve a reward of 40, and it is also the only model that achieves a reward value of 42 or more.
Compared to other models, it maintains a slight advantage.
Notably, the DQN model is the worst-performing model and requires a longer time to converge.
At the same time, PPO slightly outperforms the A2C algorithm in terms of reward and blocking rate.
\autoref{RMSA-BlockingRate} further illustrates the performance of the various algorithms mentioned above in terms of blocking rate.
By observing the two figures, we can conclude that higher rewards mean lower blocking rates.
Consistent with the results in \autoref{RMSA}, OptiGAI achieves a better blocking rate at the beginning of training, while PPO and A2C achieve similar performance after 1000 training steps.
DQN shows a slightly higher blocking probability, indicating the need for further training and hyperparameter adjustment.
The simulation results in the RMSA environment indicate that compared to other advanced DRL models, OpticGAI can first converge and maintain a slight advantage in blocking rate.
The results of AIGC-enhanced optimization decisions provide a promising future direction for optical network optimization.

\section{Discussion and Challenges}
\textbf{Exploring more Generative AI methods.}
This work takes a diffusion model-based policy network as a case study to show the feasibility of OpticGAI. 
Beyond diffusion models, other generative models, such as LLMs, have also been extensively validated for their robust generative capabilities.
Given this backdrop, integrating other advanced generative models into reinforcement learning frameworks holds significant promise for enhancing performance in network optimization tasks.
However, the motivation behind this integration and the specific implementation details warrant further elaboration and exploration. 
Key aspects that require deeper investigation include:
(1) Technically speaking, can all types of generative models be effectively incorporated into network optimization?
This question demands a thorough analysis of the compatibility between different generative models and the specific requirements of optical network tasks.
(2) What tangible benefits does the introduction of generative models bring to real-world network optimization scenarios?
This involves assessing improvements in efficiency and scalability when these models are deployed.

\textbf{Generalization capabilities.} 
A pivotal concern in deploying generative AI-aided DRL systems in optical networks is their ability to generalize across varied scenarios. 
These systems must demonstrate robustness not only across different traffic matrices, which may fluctuate significantly in volume and patterns, but also across diverse network topologies that can range from simple to complex configurations~\cite{zhang2024resource}. 
Such generalization is crucial to ensure that OpticGAI-inspired methods remain effective and efficient in real-world applications, where conditions are rarely uniform or predictable. 
Addressing this challenge requires rigorous testing and possibly the development of more sophisticated training methodologies that can teach these models to adapt to a wider range of conditions without compromising performance.
Furthermore, further system experiments need to cover real-world data instead of the simulated environments to strengthen the validation and applicability of the OpticGAI paradigm. 

\textbf{Explainability.}
The complexity of DRL models, especially when augmented by generative AI techniques such as diffusion models, often results in a "black-box" nature, where the decision-making processes are opaque and difficult to interpret. 
This lack of transparency can be a significant barrier, particularly in critical systems like optical networks where stakeholders require clear and comprehensible explanations of how decisions are made~\cite{li2024trustworthy}.
Enhancing the explainability of these models is crucial, not only for building trust and acceptance among users but also for facilitating the diagnosis and rectification of errors in the system's operations.
Explainability ensures that the decisions made by complex models can be understood, thereby increasing confidence in the stability of communication systems, which is especially important in optical network-supported low-latency transmission.
Strategies might include developing new tools and techniques for visualizing the decision-making paths and training models to generate more interpretable decisions.

\section{Related Works}
The resource allocation problem of optical networks can be dated back to the early 1990's~\cite{bala1991algorithms, ramaswami1995routing}.
Researchers continue to explore fast and scalable algorithms to address these challenges, particularly as representative problems like RWA are NP-hard.
There remains a persistent need for innovative solutions.
Among the algorithms to solve the resource allocation problem of optical networks, reinforcement learning has been increasingly adopted in recent years due to its potential to adapt to dynamic network conditions and its high generalization, thereby offering improved efficiency and scalability in optical network management.

Recent advancements in DRL have shown significant potential for optimizing network operations in optical networks.
DRL has been utilized to learn and improve heuristic algorithms for dynamic and static routing challenges such as RMSA and RWA. 
For instance, DRL techniques have been applied to optimize the OSPF weights to manage the minimum maximum link load efficiently in the network~\cite{Bernardez2021machine}.
Similarly, other heuristic algorithms have been developed for RMSA in optical networks, with adaptations to address the challenges associated with sparse rewards~\cite{Chen2018deep, cicco2022on}.

Moreover, a variety of DRL algorithms have been explored to further enhance optical network management.
Specifically, DQN integrated with Graph Neural Networks has been employed to optimize routing in optical transport networks~\cite{almasan2019deep}.
A2C and A3C methods have been used to learn heuristic algorithms for service provisioning and RMSA in EONs, respectively~\cite{li2020deepcoop, Chen2018deep}. 
MARL approach has been applied to tackle complex inter-domain RMSA tasks, showcasing the potential for distributed problem-solving in network scenarios~\cite{chen2019building}.
These developments highlight the breadth of DRL applications in optical network optimization, demonstrating its effectiveness in adapting to dynamic network conditions and enhancing overall network efficiency and scalability.

\section{Conclusion}
OpticGAI is a promising AI-generated policy paradigm in optical network optimization, which effectively handles resource-constrained and multi-type complex optical network tasks.
It allows for the automated generation of policy decision models, without the need for pre-designed network structures.
This work explores the feasibility and advantages of the OpticGAI paradigm in resource-constrained and multi-type optical network tasks.
Future work will focus on designing more universal AIGC-enhanced policy designs to obtain more flexible and efficient solutions and develop a universal framework suitable for various AIGC models.

\begin{acks}
This work was supported by the National Natural Science Foundation of China under Grant 62202302 and U20B2048.
We thank Qiaolun Zhang from Politecnico di Milano for his suggestions. 
\end{acks}

\bibliographystyle{ACM-Reference-Format}
\bibliography{sample-base}

\end{document}